\documentclass[aps,prl,reprint,superscriptaddress,showpacs]{revtex4-1}

\pdfoutput=1

\usepackage{graphicx}             
\usepackage{amsmath}              
\usepackage{amssymb}              
\usepackage{braket}              
\usepackage[load=prefixed]{siunitx}
\usepackage{color}	
\usepackage[english]{babel}
\usepackage{ifthen}
\usepackage[hidelinks]{hyperref}
\usepackage{ulem}
\usepackage{float}
\usepackage{caption}
\usepackage{appendix}

\newcommand{\figref}[1]{\mbox{Fig. \ref{#1}}}

\begin{document}
\title{Suppression of the impurity-induced local magnetism by the opening of a spin pseudogap in Ni-doped Sr$_2$CuO$_3$}
\date{\today}
\author{Yannic Utz}
\affiliation{IFW-Dresden, Institute for Solid State Research, PF 270116, 01171 Dresden, Germany}
\author{Franziska Hammerath}
\affiliation{IFW-Dresden, Institute for Solid State Research, PF 270116, 01171 Dresden, Germany}
\affiliation{Institute for Solid State Physics, Dresden Technical University, TU-Dresden, 01062 Dresden, Germany}
\author{Satoshi Nishimoto}
\affiliation{IFW-Dresden, Institute for Solid State Research, PF 270116, 01171 Dresden, Germany}
\author{Christian Hess}
\affiliation{IFW-Dresden, Institute for Solid State Research, PF 270116, 01171 Dresden, Germany}
\author{Neela Sekhar Beesetty}
\affiliation{SP2M-ICMMO UMR-CNRS 8182, Universit\'{e} Paris-Sud, 91405 Orsay Cedex, France}
\author{Romuald Saint-Martin}
\affiliation{SP2M-ICMMO UMR-CNRS 8182, Universit\'{e} Paris-Sud, 91405 Orsay Cedex, France}
\author{Alexandre Revcolevschi}
\affiliation{SP2M-ICMMO UMR-CNRS 8182, Universit\'{e} Paris-Sud, 91405 Orsay Cedex, France}
\author{Bernd B\"uchner}
\affiliation{IFW-Dresden, Institute for Solid State Research, PF 270116, 01171 Dresden, Germany}
\affiliation{Institute for Solid State Physics, Dresden Technical University, TU-Dresden, 01062 Dresden, Germany}
\author{Hans-Joachim Grafe}
\affiliation{IFW-Dresden, Institute for Solid State Research, PF 270116, 01171 Dresden, Germany}

\begin{abstract}
The \mbox{$S=1/2$} antiferromagnetic Heisenberg spin chain compound Sr$_2$CuO$_3$ doped with $\SI{1}{\percent}$ and $\SI{2}{\percent}$ of Ni impurities has been studied by means of $^{63}$Cu nuclear magnetic resonance. A strong decrease of the spin-lattice relaxation rate $T_1^{-1}$ at low temperatures points toward a spin gap, while a stretching exponent $\lambda < 1$ and a frequency dependence of $T_1^{-1}$ indicate that this spin gap varies spatially and should rather be characterized as a spin pseudogap. 
The magnitude of the spin pseudogap scales with doping level.
Our results therefore evidence the finite-size character of this phenomenon.
Moreover, an unusual narrowing of the low temperature NMR lines reveals the suppression of the impurity-induced staggered paramagnetic response with increasing doping level.
\end{abstract}


\maketitle

The one-dimensional (1D) \mbox{$S=1/2$} antiferromagnetic Heisenberg model can be called the harmonic oscillator of quantum magnetism.
Its integrability makes it exactly solvable and therefore it is often used as an archetype for low-dimensional quantum magnets in theory.  
Despite its simplicity, it shows a very unusual behavior. 
The ground state of this model is an example of an highly entangled many-body quantum state, which is characterized by a lack of long-range order even at absolute zero. Its elementary excitations are exotic quasiparticle excitations with fractional quantum numbers, the $S=1/2$ spinons, which can be excited with infinitely low energy, i. e. the excitation spectrum has no energy gap to the ground state \cite{Auerbach1994,Schollwoeck2004,Lieb1961,Cloizeaux1962}. 
Regardless of its fundamentally important role, it is difficult to find realizations of this model in nature. 
Small perturbations induced by impurities or by interchain interactions are expected to lead to gaps in the excitation spectra or to three-dimensional (3D) long-range ordering \cite{Affleck1989,Eggert1992,Affleck1994,Kojima1997,Laukamp1998, Eggert2002}.
It is therefore important to perform clear-cut experiments which explicitly address these perturbations in a controlled manner.
The investigation of the staggered paramagnetic response around intrachain impurities has been proposed to be a valuable tool for this purpose \cite{Alloul2009}.

The cuprate compound Sr$_2$CuO$_3$ is known to be among the best realizations of the 1D \mbox{$S=1/2$} antiferromagnetic Heisenberg model. There, the chains are realized by corner sharing CuO$_4$ plaquettes with \mbox{$S=1/2$} on the copper site, which are mainly interacting along one crystallographic axis with a large exchange coupling of about $J \sim \SI{2000}{K}$ \cite{Motoyama1996}.
Weak static magnetism occurs only below $T_N = \SI{5.4}{K}$ \cite{Keren1993, Kojima1997}, which is low compared to the much larger exchange coupling $J$. 
However, recent studies on doped variants of Sr$_2$CuO$_3$ and the closely related double chain compound SrCuO$_2$ revealed the vulnerability of the originally gapless spinon excitation spectrum \cite{Zaliznyak2004, Takigawa1996} against the influence of impurities and disorder.
$^{63}$Cu nuclear magnetic resonance and transport studies showed that doping Ca on the Sr site outside the chains breaks the integrability of the model and opens a spin gap of similar size in both compounds, which has been attributed to structural distortions and a concomitant bond disorder \cite{Hammerath2011a, Hlubek2011, Hammerath2014, Mohan2014}.
Inelastic neutron scattering disclosed a striking impact of minor concentrations of intrachain nickel impurities on the low-energy spin dynamics of the double chain compound SrCuO$_2$ \cite{Simutis2013}. 
The authors report the emergence of a spin pseudogap of the order of $\Delta\approx\SI{90}{K}$ by replacing only $\SI{1}{\percent}$ of the \mbox{$S=1/2$} copper ions with \mbox{$S=1$} nickel impurities. 
Corresponding to their interpretation the nickel spin is fully screened. Therefore, the nickel ions effectively act as S=0 impurities and basically cut the chains into segments with varying finite length $l$, 
which show finite-size spin gaps with magnitudes proportional to $1/l$ \cite{Eggert1992}.

In this article, we show results on the doping-dependent effect of nickel impurities on the single chain compound Sr$_2$CuO$_3$ based on nuclear magnetic resonance (NMR). Spin-lattice relaxation measurements reveal the opening of a spin pseudogap that scales linearly with the Ni content and therefore prove the finite-size nature of this phenomenon. However, the NMR spectra evidence a suppression of the impurity-induced staggered paramagnetic response with increasing impurity content, in stark contrast to what has been observed so far in gapped low dimensional spin systems \cite{Alloul2009, Fujiwara1998, Ohsugi1999, Bobroff2009a, Casola2010, Hase1993a}.  

The measurements were performed on high purity single crystals of Sr$_2$Cu$_{1-x}$Ni$_{x}$O$_3$ ($x= 0.01$ and $0.02$, labeled Ni1 and Ni2 hereafter).
The samples were prepared using the travelling solvent floating zone (TSFZ) method, since this compound undergoes a peritectic-type decomposition upon cooling from the melt \cite{Revcolevschi1999}. The used starting powders of SrCO$_3$, NiO and CuO were of \SI{99.99}{\percent} purity. The crystals were grown at a growth rate of \SI{1}{mm/h} under flowing oxygen atmosphere (\SI{50}{ml/min}). A solvent pellet of composition $\SI{63}{\percent}$CuO-\SI{37}{\percent}SrO was used to initiate the crystal growth experiment.
The crystals cleave readily along ($h, 0, 0$). 
The high quality of the crystals has been checked by X-ray diffraction (phase determination) and energy-dispersive X-ray spectroscopy (chemical composition) measurements.

The $^{63}$Cu NMR spectra have been measured with the standard Hahn spin echo method at a fixed frequency of $\SI{80}{MHz}$ by sweeping the external magnetic field $H$ and integrating the echo.
At room temperature, the  NMR spectra of the ($I=3/2$) $^{63}$Cu nuclei  consist of three narrow quadrupolar split lines with satellites which are only slightly affected by quadrupolar broadening \footnote{The linewidths (FWHM) of the $^{63}$Cu spectra at $T=\SI{300}{K}$ and $\SI{80}{MHz}$ are $\Delta B_{ml}=\SI{3 \pm 1}{mT}$ and $\Delta B_{sat}=\SI{6\pm 1}{mT}$ for the main line and the satellites of Ni1, and $\Delta B_{ml}=\SI{3 \pm 1}{mT}$ and $\Delta B_{sat}=\SI{10 \pm 2}{mT}$ for the main line and the satellites of Ni2. See also appendices.} that results from the structural disorder mainly induced by the Ni-dopants.

The $^{63}$Cu NMR spin-lattice relaxation rate $T_1^{-1}$ has been measured by the inversion recovery method on one of the quadrupolar split satellite lines for both doping levels. 
All $T_1$ measurements were performed in magnetic fields close to $\SI{7}{T}$ \footnote{$T_1$ measurements on Ni1 have been performed in fields of $\mu_0 H = \SI{7.0493}{T}$ and $\mu_0 H = \SI{7.3064}{T}$, while all measurements on Ni2 have been performed at $\mu_0 H = \SI{7.0488}{T}$.} with $H$ parallel to the crystallographic \textit{a} axis.
Accurate alignment of the samples has been achieved by utilizing the angle dependence of the second order quadrupolar shift of the $^{63}$Cu main line.
The recovery curves of the nuclear magnetization have been fit to the standard function for magnetic relaxation of $I=3/2$ nuclei measured on a satellite transition \cite{McDowell1995, Johnston2006}:
\begin{flalign}
M_z(t)=M_0 \Big[ 1 - f \Big( 0.4 e^{-(6t/T_1)^\lambda} &+ 0.5 e^{-(3t/T_1)^\lambda}&  \label{eq:recoberySat}\\
     \mathrel{\phantom{=}}  &+ 0.1e^{-(t/T_1)^\lambda} \Big) \Big]   . \nonumber
\end{flalign}
$M_0$ is the equilibrium value of the nuclear magnetization, $f$ is ideally $2$ for a complete inversion and the stretching exponent $\lambda \leq 1$ accounts for a distribution of spin-lattice relaxation rates around a characteristic value $T_1^{-1}$. For high temperatures, $\lambda \approx 1$ shows that there is a unique spin-lattice relaxation rate $T_1^{-1}$. For lower temperatures, $\lambda < 1$ indicates a spatial distribution of nuclei with different spin-lattice relaxation rates.
The ratio of $T_1$ of the two Cu isotopes indicates purely magnetic relaxation over the whole temperature range.

\begin{figure}
\includegraphics[width=8cm]{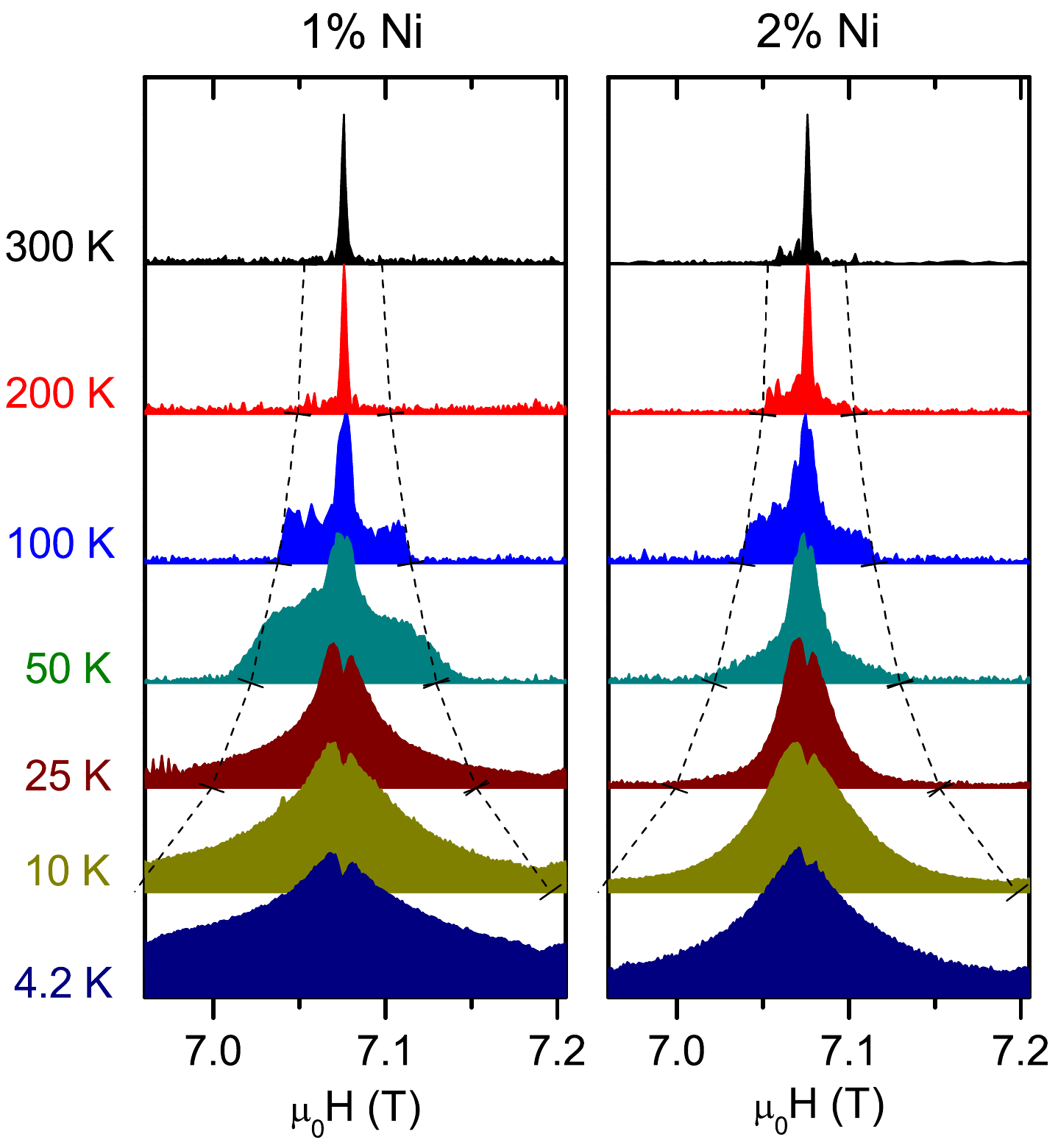}
\caption{$^{63}$Cu NMR main line of Sr$_2$CuO$_3$ doped with $\SI{1}{\percent}$ and $\SI{2}{\percent}$ of nickel at various temperatures obtained at a fixed frequency of $\SI{80}{MHz}$ by varying the external field. The intensity of the resonance lines is normalized to their maxima. The dashed lines indicate the expected $1/\sqrt{T}$ behavior of the shoulder feature and has been fit to its width at 100K.}
\label{fig:spectra}
\end{figure}

We will start our discussion with the resulting NMR spectra.
\figref{fig:spectra} shows the $^{63}$Cu main line for both doping levels and various temperatures. Upon lowering the temperature, the spectra show a pronounced broadening. They develop shoulder structures and a splitting of the central peak.  Both, the central line and the satellites, are equally affected (see appendices), which indicates that the broadening is of magnetic origin. This means that the spectra can be seen as a histogram of the distribution of local magnetic fields. It is well known that such a magnetic broadening within antiferromagnetically correlated systems can be attributed to the presence of clouds of field-induced staggered polarization around impurities \cite{Alloul2009, Fujiwara1998, Ohsugi1999, Bobroff2009a, Casola2010, Hase1993a}. The shape and extension of such a local alternating magnetization (LAM) depends very much on the nature of the underlying spin system and the nature of its coupling to the impurity spin.

A LAM has been observed previously in undoped Sr$_2$CuO$_3$, where it has been explained by open chain ends due to excess oxygen \cite{Takigawa1997a,Boucher2000,Sirker2009}. These chain ends break the translational invariance of the spin chain  and lead to a local alternating susceptibility  [$\chi_\mathrm{alt}(x)$], which gives rise to a LAM in a magnetic field.
It could be modeled based on the assumption of semi-infinite chains \cite{Eggert1995}, which predicts a LAM with a maximum at a certain distance $l=\num{0.48}J/T$ \cite{Takigawa1997a} from the impurity and an exponential decay for larger distances. Upon lowering the temperature, the maximum shifts further into the chain and increases with $\chi_\mathrm{alt,max} \propto 1/\sqrt{T}$.
In the NMR spectra, this causes a broad background with sharp edges, which broadens with decreasing temperature corresponding to $\Delta H \propto 1/\sqrt{T}$ independent of the amount of chain breaks \cite{Eggert1995, Takigawa1997a}. The intensity of the background should increase with decreasing temperature. 

We can identify the shoulder features as this broad background. 
Owing to a larger impurity concentration than in the undoped compound, the features are already well developed at $\SI{100}{K}$ for Ni1. For Ni2, one can already observe shoulder features at $\SI{200}{K}$.
The dashed lines in \figref{fig:spectra} indicate the expected $1/\sqrt{T}$ behavior \footnote{The $1/\sqrt(T)$ behavior was fitted to the width $\Delta H$ of the shoulder feature at $\SI{100}{K}$ and corresponds to $\sqrt(T)(\Delta H / 2H_0) = 0.0544$, which is an agreement with the results of \cite{Takigawa1997a}. }. The line shape clearly follows this trend.
However, at lower temperatures the shoulder structure is smeared out. The onset of the smearing depends on the doping level. It sets in at higher temperatures for Ni2 than for Ni1.
Moreover, one can see that the resonance lines at low temperatures (below $\SI{100}{K}$) are narrower for Ni2 than for Ni1.
This smearing of the shoulder structures and also the narrowing of the low-temperature resonance lines with increasing doping level are surprising and cannot be explained by the simple approach to $\chi_{\mathrm{alt}}(x)$ mentioned above. 

At low enough temperatures, the LAM is predicted to extend over the whole chain segment and the assumption of the semi-infinite chain is no longer valid \cite{Eggert1995, Laukamp1998, Alloul2009, Sirker2009}. Due to the high impurity content, the finite size of the chain segments cannot be neglected in the investigated temperature range. However, it is not expected to result in a reduced width as compared to the $1/\sqrt{T}$ behavior or even a disappearance of the shoulder features.
The deviations might be connected to the screening of the Ni spin. Since Ni is a magnetic impurity, an additional screening cloud is expected to contribute to the LAM \cite{Rommer2000, Eggert1992}. Moreover, susceptibility measurements on Sr$_2$Cu$_{0.99}$Ni$_{0.01}$O$_3$ show that the nickel spin is in fact screened \cite{Karmakar2015}.
But such a screening cloud should rather enhance the width of the NMR line than suppress it. 
The suppression of the linewidth might also be connected to the results on the low-energy excitation spectrum, which will be discussed in the following.

\begin{figure}
\includegraphics[width=8cm]{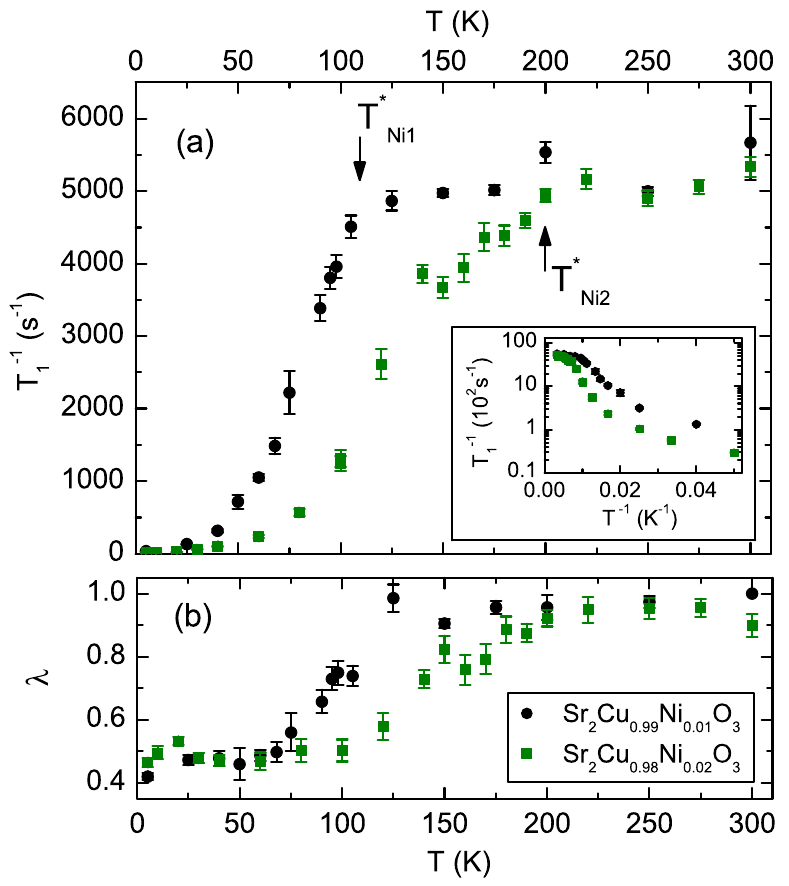}
\caption{$^{63}$Cu NMR spin-lattice relaxation rate $T^{-1}_1$ (a) and stretching exponent $\lambda$ (b) of Sr$_2$Cu$_{0.99}$Ni$_{0.01}$O$_3$ (black circles) and Sr$_2$Cu$_{0.98}$Ni$_{0.02}$O$_3$ (green squares) measured at the center of the high field satellites. Inset of (a): Arrhenius plot of the spin-lattice relaxation rates.}
\label{fig:T1}
\end{figure}

\mbox{\figref{fig:T1}(a)} shows the temperature dependence of $T_1^{-1}$ measured at the center of the resonance lines for Ni1 and Ni2. At high temperatures, $T_1^{-1}$ is temperature independent for both dopings, as it is theoretically expected for antiferromagnetic \mbox{$S=1/2$} Heisenberg chains \cite{Sachdev1994, Sandvik1995} and as it has been experimentally verified for the parent compound \cite{Takigawa1996}.
Below a certain crossover temperature, which is  $T^\ast_{\mathrm{Ni}1}\approx\SI{110}{K}$ for Ni1 and $T^\ast_{\mathrm{Ni}2}\approx\SI{200}{K}$ for Ni2, $T_1^{-1}$ shows a strong decrease by two orders of magnitude toward low temperatures.
The decrease of $T_1^{-1}$ is accompanied by a decrease of the stretching exponent $\lambda$ [see \mbox{\figref{fig:T1}(b)}] and thus by a growing spatial distribution of spin-lattice relaxation rates, which
levels off at lower temperatures.

Due to the hyperfine coupling $A_\perp$ between nuclei and electrons, $T_1^{-1}$ measures the imaginary part of the dynamic spin susceptibility $\chi^{\prime\prime}$ of the electronic spin system at the NMR frequency.
For pure magnetic relaxation, it is given by 
\begin{equation}
T_1^{-1} \propto T \sum_{\vec{q}}A^2_\perp (\vec{q},\omega ) \frac{\chi^{\prime\prime}(\vec{q},\omega)}{\omega} \, .
\end{equation}
On a more intuitive level, the relaxation mechanism can be described as the scattering of thermally excited spinons by the copper nuclei \cite{Magishi1998}.

Thus, the decrease in spin-lattice relaxation rates clearly indicates the depletion of low-lying states in the spin excitation spectrum, and therefore points toward a spin gap. However, the distribution of spin-lattice relaxation rates, as indicated by $\lambda < 1$, implies that this spin gap varies spatially and should rather be characterized as a spin pseudogap \footnote{We also tried to fit the recovery curves assuming a distribution of $T_1$ due to a distribution of gaps due to a distribution of chain length. These attempts failed. See appendices for details.}.

Usually, the magnitude of a spin gap is estimated by fitting the temperature dependence of $T_1^{-1}$ to an activated behavior \cite{Hammerath2011a,Takigawa1998,Ishida1994,Ohama1997, Imai1998} and using the activation energy as an estimate for the spin gap.
However, in our case, the spin-lattice relaxation rates do not decrease exponentially [see the inset of \mbox{\figref{fig:T1}(a)}]. This can be attributed to the spatial distribution of spin gaps, because the fast relaxation stemming from nuclei exposed to small gaps will dominate the recovery process at low temperatures. We use the crossover temperature $T^\ast$ as an estimate for the average gap energy. $T^\ast_{\mathrm{Ni}2}$ is about twice as large as $T^\ast_{\mathrm{Ni}1}$. Therefore, we conclude that the spin pseudogap is proportional to the doping level. 
This is in agreement with the assumption that the individual chain segments show gaps $\Delta \propto 1/l$ and thus evidences the finite-size character of the spin pseudogap.
The value of $T^\ast_{\mathrm{Ni}1} \approx \SI{110}{K}$ is close to the reported spin pseudogap $\Delta\approx\SI{90}{K}$ \cite{Simutis2013} of the double chain compound doped with $\SI{1}{\percent}$ of nickel, which suggests that the double chain structure is not crucial to the gapping mechanism, similar to what has been observed in the Ca-doped variants of SrCuO$_2$ and Sr$_2$CuO$_3$ \cite{Hammerath2011a, Hammerath2014}.

\begin{figure}
\includegraphics[width=8cm]{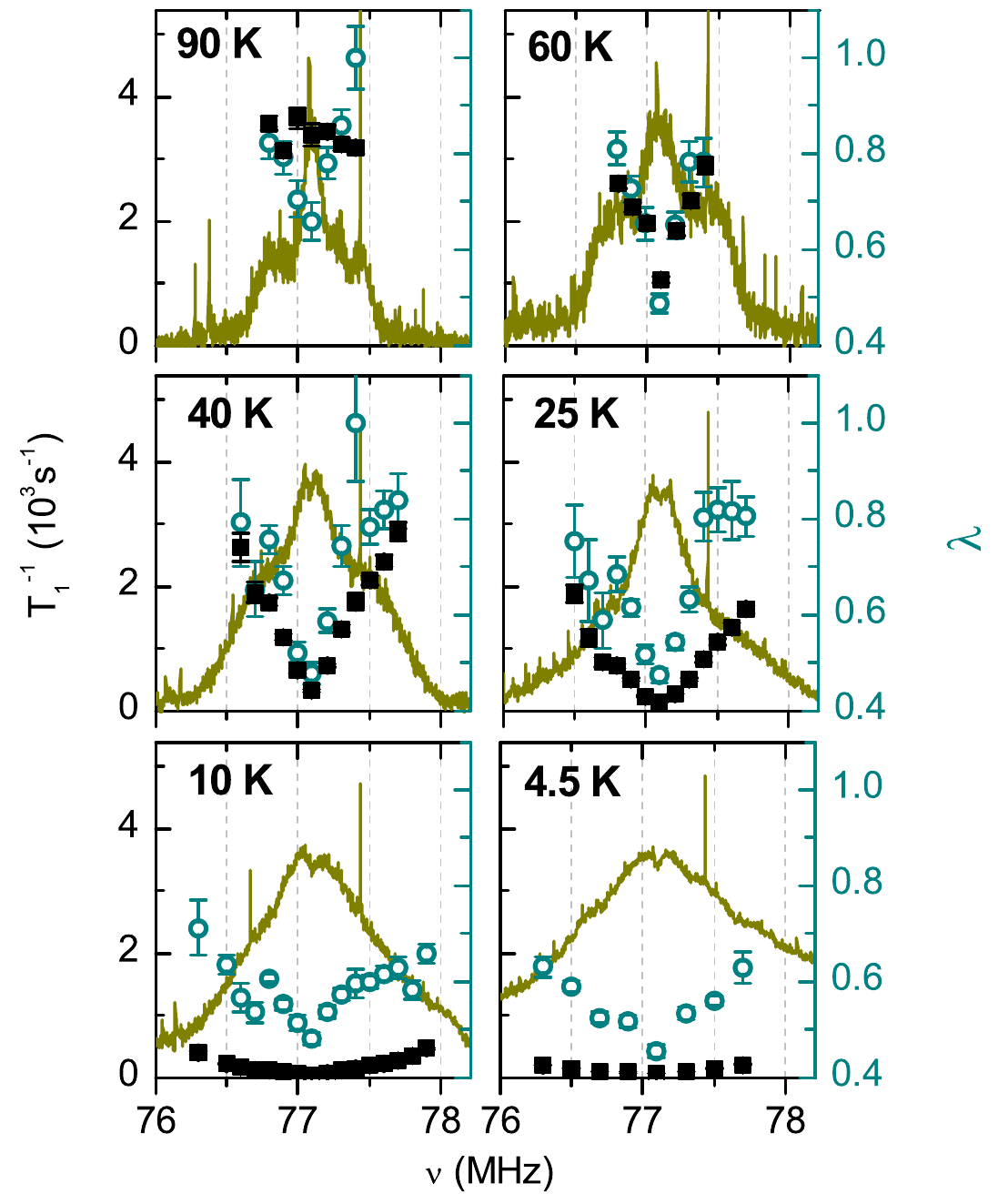}
\caption{Spin-lattice relaxation rate $T^{-1}_1$ (solid squares) and stretching exponent $\lambda$ (open circles) of Sr$_2$Cu$_{0.99}$Ni$_{0.01}$O$_3$ measured at different positions within  the $^{63}$Cu high field satellite (solid lines) at various temperatures in a magnetic field of $\mu_0 H = \SI{7.0493}{T}$ with $H||a$. The intensity of the resonance lines is scaled to arbitrary units.}
\label{fig:freqdepT1}
\end{figure}

Having established the proportionality between the doping level and the average gap energy, we wanted to gain further knowledge about the spatial variation of spin gaps.
Therefore, we investigated the frequency dependence of $T_1^{-1}$ within the broad resonance lines.
\figref{fig:freqdepT1} shows spin-lattice relaxation rates $T_1^{-1}$ and stretching exponents $\lambda$ measured at different positions within the high field satellite of Ni1 \footnote{For the frequency-dependent measurements of the spin-lattice relaxation rate, the inversion recovery method with $\ang{90}$-pulses ranging from \SI{3}{\micro\second} to \SI{4}{\micro\second} was used. The recovery curves were obtained by integrating the echo intensity and  therefore selectively show the relaxation of the nuclei being in resonance with the used frequency.}. The spectra themselves are also shown for guidance. In contrast to the spectra shown before, these spectra were obtained at a fixed field $\mu_0H=\SI{7.0493}{T}$ by sweeping the frequency and adding the Fourier transforms of the echo signals (frequency step and sum method \cite{Clark1995}).
While $T_1^{-1}$ is frequency-independent for $T \geq \SI{90}{K}$, it shows a strong frequency dependence at lower temperatures. Spin-lattice relaxation rates at all positions decrease toward low temperatures, but the decrease of $T_1^{-1}$ is less pronounced for larger distances to the center of the resonance line. The frequency dependence of $T_1^{-1}$ is accompanied by a frequency dependence of the stretching exponent $\lambda$, which sets in already  at $T=\SI{90}{K}$. $\lambda$ is minimal at the center and larger at the outer parts of the resonance lines. This indicates that Cu nuclei which contribute to the outer parts of the resonance lines probe a narrow distribution of small spin gaps, while Cu nuclei contributing to the center of the resonance lines probe a broad distribution of large and small spin gaps. 
As NMR is a spectroscopic method, we cannot distinguish if the gap differs only between chain segments of different length or if it also varies within individual chain segments. If the shape of the LAM in real space would be known, such a distinction could become possible.
Besides, we cannot exclude that additional impurity-induced spin fluctuations the near chain ends lead to a variation of $T_1^{-1}$ within single chain segments and therefore contribute in combination with the LAM to the frequency dependence of $T_1^{-1}$. Such spin fluctuations may also enhance the deviation of $T_1^{-1}$ from exponential behavior at low T [see the inset of \figref{fig:T1}(a)].
However, we can state that the largest gaps are measured at the center of the resonance lines and therefore by nuclei not exposed to the LAM. This might be the key to understand the suppression of the LAM at low T and the frequency dependence of $T_1^{-1}$. It suggests that the LAM is suppressed by the gap.
This is supported by the fact that the gap increases with increasing impurity concentration, while the spectral broadening is reduced.
In view of this considerable suppression of local magnetism, it is not surprising that the ordering temperature is strongly reduced too. We did not find any signature of magnetic ordering down to $\SI{4.2}{K}$ for both dopings. This is in agreement with recent measurements of the susceptibility and the specific heat of Sr$_2$Cu$_{0.99}$Ni$_{0.01}$O$_3$, which do not show any transition down to \SI{2}{K} \cite{Karmakar2015}.

In summary, our $^{63}$Cu NMR measurements on single crystals of the $S=1/2$ spin chain Sr$_2$Cu$_{1-x}$Ni$_{x}$O$_3$ ($x=0.01$, $0.02$) show a strong impact of minor concentrations of nickel on the low-energy spin dynamics and the local susceptibility of the spin chains.
We find a doping-dependent spin pseudogap behavior,  which evidences the finite-size character of this phenomenon.
The NMR spectra show a local alternating magnetization around the nickel impurities. Its suppression at low temperatures, the variation of $T_1^{-1}$ within the broad resonance lines and the strong reduction of the ordering temperature are most probably consequences of the spin pseudogap, which reduces low-energy antiferromagnetic fluctuations.

\begin{acknowledgments}
The authors thank S.-L. Drechsler and L. Hozoi for discussion.
This work has been supported by the European Commisssion through the LOTHERM project (Project No. PITN-GA-2009-238475) and by the Deutsche Forschungsgemeinschaft (DFG) through Grant No. GR3330/4-1, through the D-A-CH project No. HE3439/12 and through the Sonderforschungsbereich (SFB) No. 1143. 
\end{acknowledgments}

%

\appendix
\appendixpage

\section{Attempt to model the distribution of $T_1^{-1}$ to fit the recovery curves}
As demonstrated in the main article, the measured recovery curves of the nuclear magnetization at low temperatures are stretched due to a distribution of spin lattice relaxation rates $T_1^{-1}$.
We took this distribution into account by fitting the phenomenological stretched exponential function to the measured recovery curves. However, this approach just gives a rough idea about the width of the distribution in terms of the stretching exponent $\lambda$. It is also possible to convolute the recovery function $M_z(t, T_1)$ with a chosen probability distribution function of $T_1$. In accordance with the approach described by \cite{Simutis2013}, we modeled the distribution of $T_1$ as result of the distribution of gaps due to a given distribution of chain length. Following \cite{Simutis2013}, the probability to find a copper atom inside a non-interrupted segment of length $L$ for a totally random distribution of defects with a concentration $x$ is given by
\begin{equation}
P_L = \frac{L \cdot x^2(1-x)^L}{\sum_{i=1}^\infty L \cdot x^2(1-x)^L} = x^2 L (1-x)^{L-1} \, .
\end{equation}
A segment of length $L$ has an energy gap $\Delta_L = \Delta_0/L$ with $\Delta_0 = 3.65 J$ \cite{Simutis2013, Eggert1992}. Assuming an activated behavior, the temperature dependence of $T_1$ for copper nuclei within a chain segment of length $L$ is given by
\begin{equation}
T_1(T,L) = T_{1,\infty} \exp(\Delta_L/T) = T_{1,\infty} \exp(\Delta_0/L/T) \, . \label{eq:activatedBehavior}
\end{equation}
$T_{1,\infty}$ labels the high temperature limit, which is independent of thegap magnitude.
The recovery function for magnetic relaxation of $I=3/2$ nuclei measured on a satellite transition with a unique 
spin-lattice relaxation time $T_1$ is \cite{McDowell1995}:
\begin{flalign}
M_z(t,T_1)=M_0 \Big[ 1 - f \Big( 0.4 e^{-6t/T_1} &+ 0.5 e^{-3t/T_1}  \label{eq:recoverySat}\\
     \mathrel{\phantom{=}}  &+ 0.1e^{-t/T_1} \Big) \Big]   . \nonumber
\end{flalign}

The recovery function based on the distribution of chain length can then be obtained as weighted sum over $L$ of Eq. (\ref{eq:recoverySat}), where $T_1$ is given by Eq. (\ref{eq:activatedBehavior}):
\begin{equation}
M_z(t) = \sum_{1}^{\infty} M_z(t,T_1(T,L)) \cdot P_L  \, .
\end{equation}

\begin{figure}[h]
\includegraphics[width=\columnwidth]{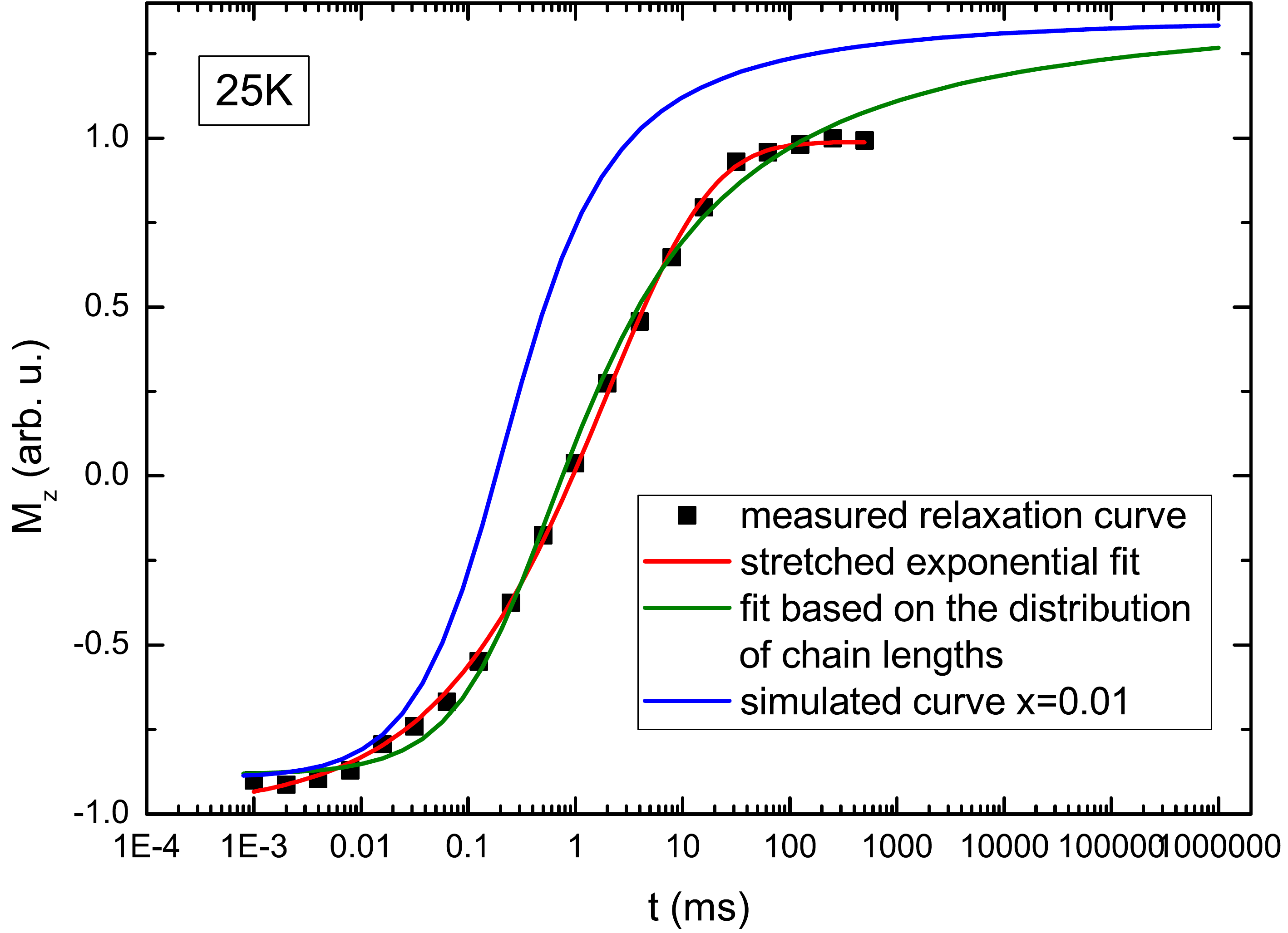}
\caption{Recovery curve of the nuclear magnetization measured at the center of the $^{63}$Cu high field satellite of Sr$_2$Cu$_{0.99}$Ni$_{0.01}$O$_3$ at $\SI{25}{K}$ in a magnetic field $\mu_0 H = \SI{7.0493}{T}$ oriented  parallel to the crystallographic \textit{a}-axis. The green line illustrates the fit with the described model. The fitting parameters are: $x=0.020 \pm 0.003$, $T_{1,\infty}=\SI{0.22 \pm 0.07}{ms}$, $M_0=1.36 \pm 0.08$ and $f=1.66 \pm 0.06$. $T=\SI{25}{K}$ and $\Delta_0=\SI{7300}{K}$
where set fixed during the fitting procedure. The blue line is simulated corresponding to the model with the same parameters as the fit except of $x=0.01$. The red line depicts a stretched fit for comparison, with $T_1= \SI{7.5 \pm 0.3}{ms}$ and $\lambda = 0.47 \pm 0.01$.    }
\label{fig:relcurve}
\end{figure}

For the fitting, we set \mbox{$\Delta_0 = 3.65 \cdot \SI{2000}{K} = \SI{7300}{K}$} and $T$ corresponding to the current temperature fixed. $x$, $T_{1,\infty}$, $M_0$, and $f$ were used as adjustable parameters.

However, it was not possible to obtain satisfactorily fits.  Fig. \ref{fig:relcurve} shows a curve measured at \SI{25}{K} and its fit as an example. One can see, that the fitting curve does not reach full recovery for very long time in contrast to the measured curve. Moreover, the resulting doping level $x=0.02$ is double as large as the nominal one. Moreover, the parameters $x$ and $T_{1,\infty}$ vary a lot over temperature, which should not be the case. Setting $x$ and $T_{1,\infty}$ to reasonable values and holding them constant during the fitting procedure resulted in even worse fit quality. 

The reason for the failed fitting could be that the original distribution of gaps is spread over the broad resonance lines, so that only excerpts of the original distribution are measured at different positions, as stated in the main article. This is supported by the differences between measured and fitted curves.
In Fig. \ref{fig:relcurve} one can see that the measured curve realizes a narrower distribution of $T_1$ than assumed by the model. This is manifested by the fact that the initial recovery of the measured curve is much slower than it should be for $x=0.01$ (see simulated curve in Fig. \ref{fig:relcurve}) and that the very slow recovery at long times it not visible in the measured data.

\section{Spectra and Linewidth}
\begin{figure*}[B]
\includegraphics[width=0.65\textwidth]{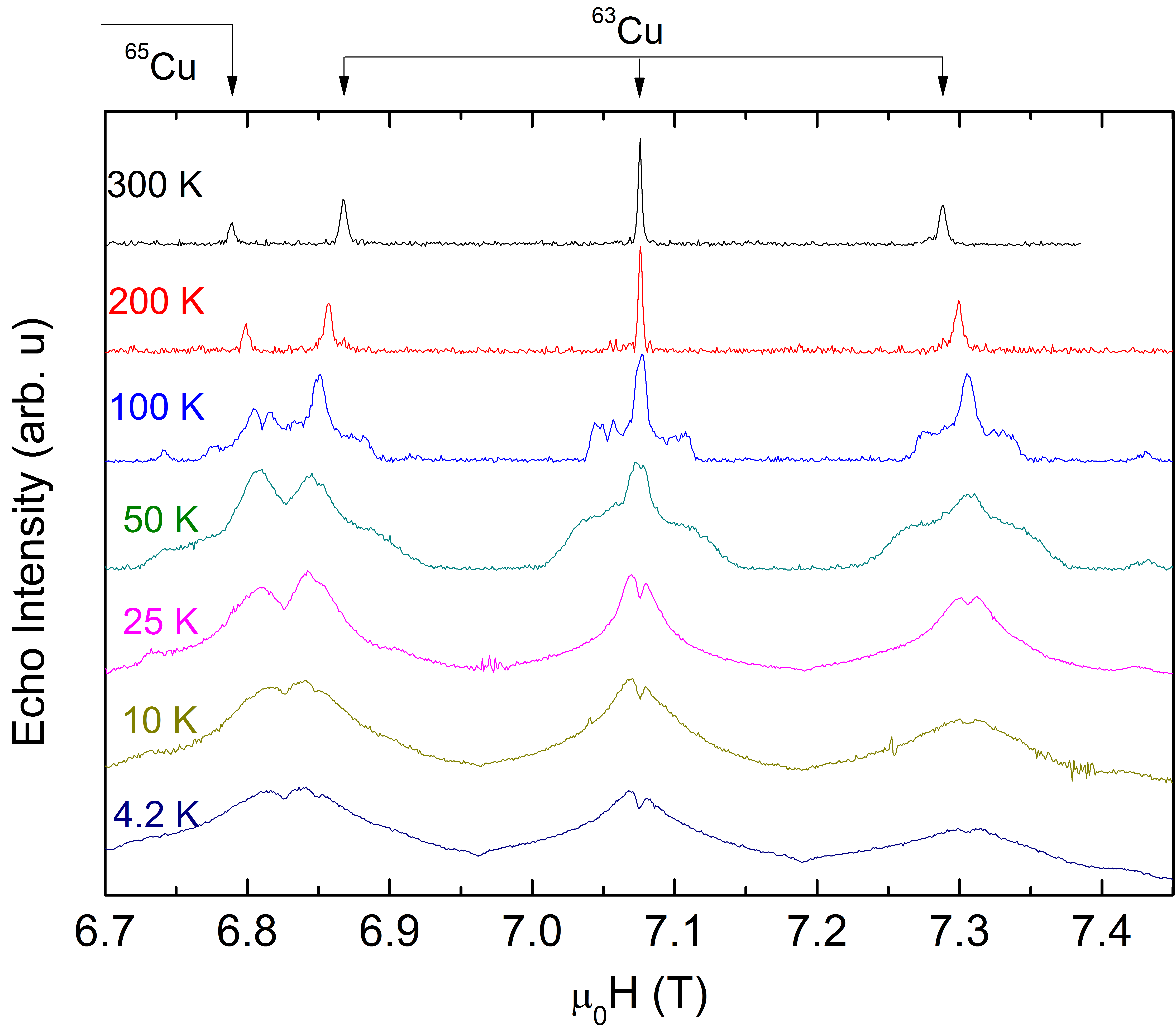}
\caption{Field-swept Cu NMR spectra of Sr$_2$Cu$_{0.99}$Ni$_{0.01}$O$_3$.  }
\label{fig:tempdepspec1Ni}
\end{figure*}

\begin{figure*}[h]
\includegraphics[width=0.65\textwidth]{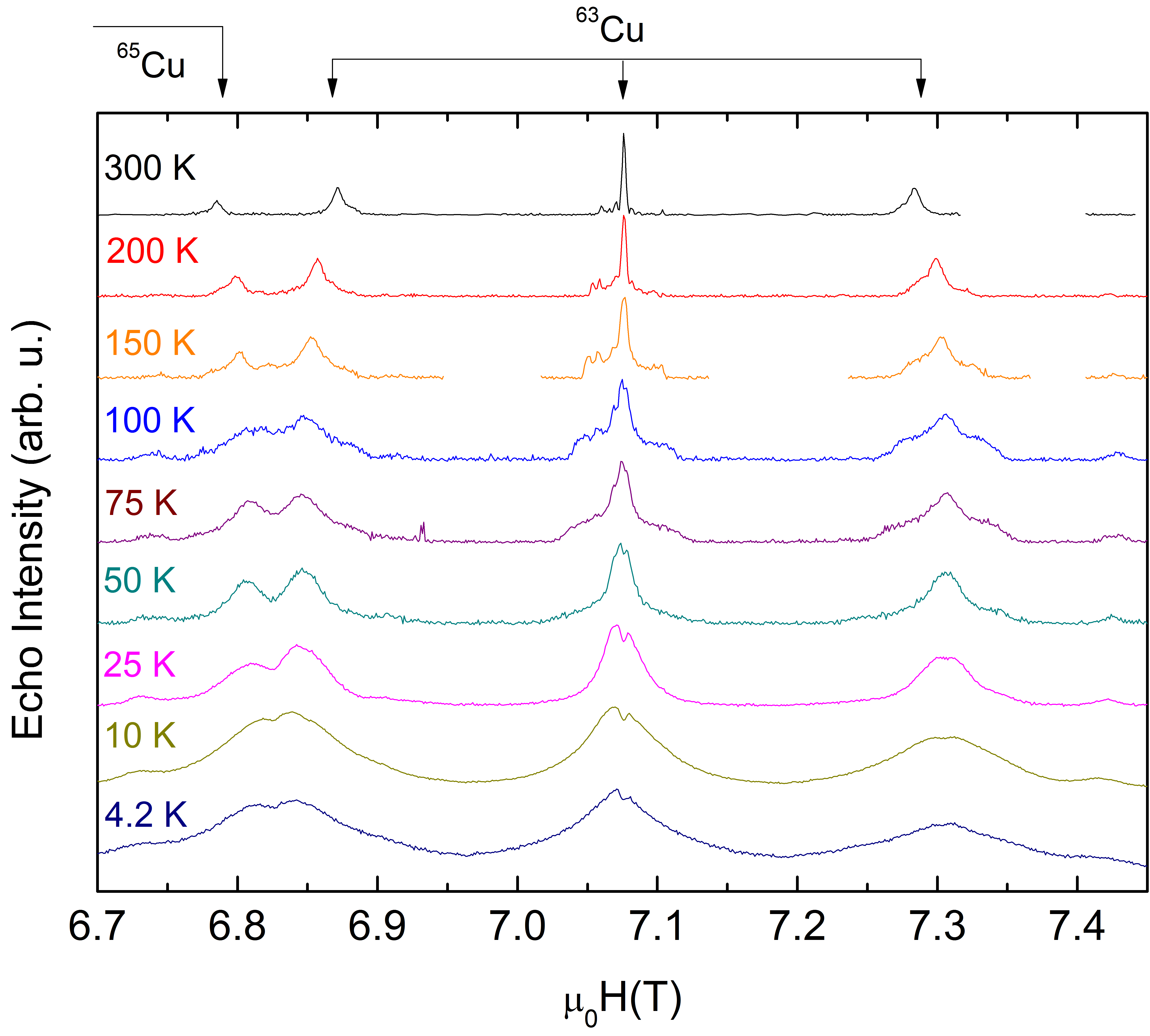}
\caption{Field-swept Cu NMR spectra of Sr$_2$Cu$_{0.98}$Ni$_{0.2}$O$_3$.}
\label{fig:tempdepspec2Ni}
\end{figure*}

\mbox{Fig. \ref{fig:tempdepspec1Ni} and \ref{fig:tempdepspec2Ni}} show field-swept Cu NMR spectra of Sr$_2$Cu$_{0.99}$Ni$_{0.01}$O$_3$ and Sr$_2$Cu$_{0.98}$Ni$_{0.2}$O$_3$ obtained with the standard Hahn spin echo method at a fixed frequency of \SI{80}{MHz}  by sweeping the magnetic field $H$ along the crystallographic \textit{a} axis and integrating the echo. They contain all three resonance lines of the $^{63}$Cu spectrum and the $^{65}$Cu high field satellite.

\begin{figure*}[h]
\includegraphics[width=0.65\textwidth]{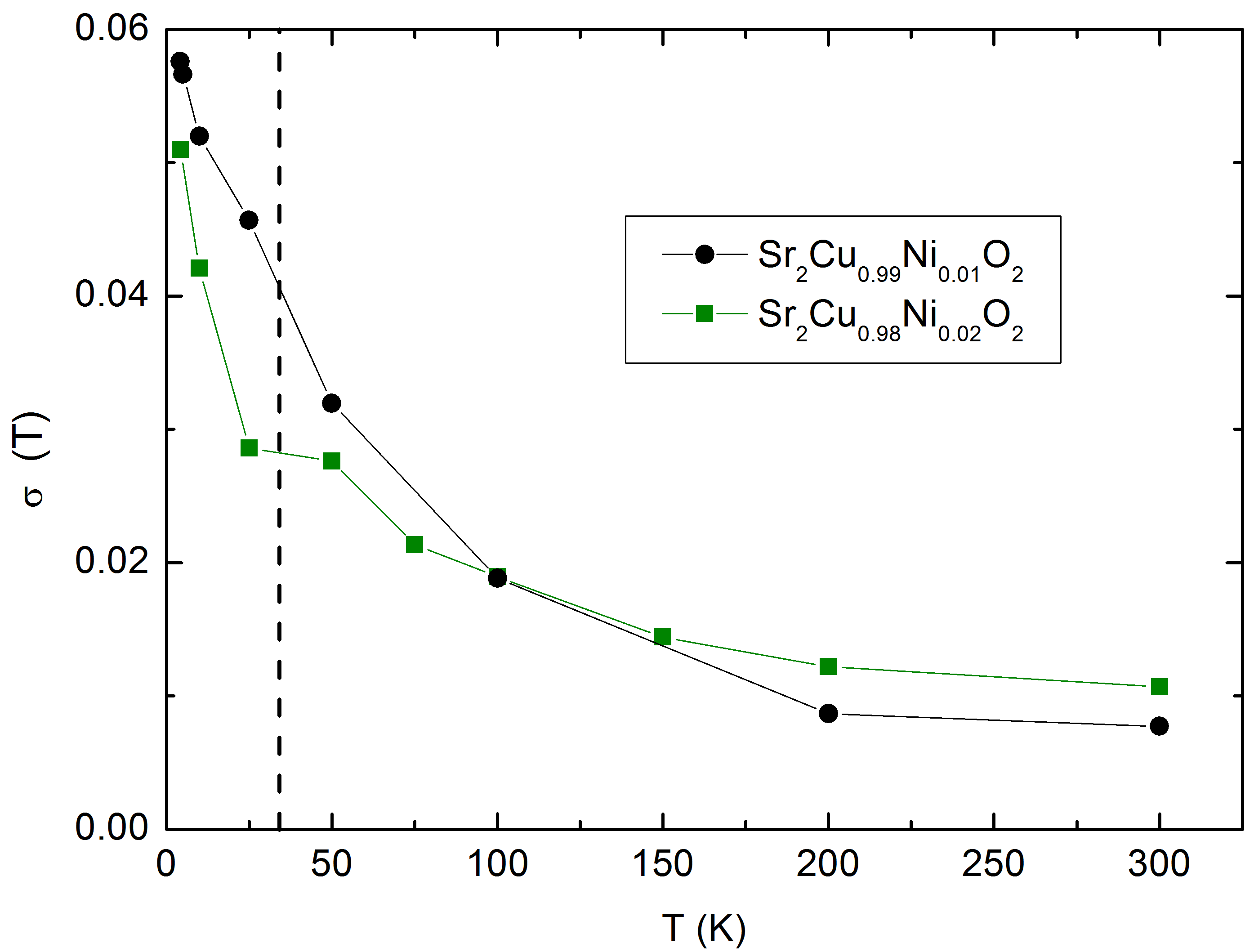}
\captionsetup{width=0.65\textwidth}
\caption{Linewidth $\sigma$ of the $^{63}$Cu main line within the spectra shown above, determined as the square root of the second moment, as a function of temperature. The solid lines are guides for the eyes. The dashed line indicates the temperature below which the second moment is not a reliable measure of the line width any more due to the overlap with the satellite lines.}
\label{fig:2ndmoment}
\end{figure*}

As the shape of the resonance lines changes drastically over temperature, we use the second moment as a measure of linewidth (see Fig. \ref{fig:2ndmoment} for the linewidth of the $^{63}$Cu main line). For temperatures $T\leq\SI{25}{K}$, the tails of the main line have an overlap with the tails of the neighboring satellite lines. In these cases, we integrated only between the two minima surrounding the mainline. This leads to incorrect values for temperatures $T\leq\SI{25}{K}$. In spite of these limitations, Fig. \ref{fig:2ndmoment} shows that the increase of the linewidth towards low temperatures is smaller for \SI{2}{\percent} Ni doping than for \SI{1}{\percent} Ni doping.

\end{document}